\documentclass{iopjournal}
\usepackage{amsmath, mathtools, natbib}

\newcommand\erf{\operatorname{erf}}

\begin{document}

\articletype{Paper} 

\title{Systematic Spectral Distortion from Digital Whitening in Radio Telescopes and Implications for 21 cm Cosmology}

\author{Ruby Byrne$^{1,2,*}$\orcid{0000-0003-4980-2736}, Larry R. D’Addario$^{1,2}$, Daniel C. Jacobs$^3$\orcid{0000-0002-0917-2269}, and Gregg Hallinan$^{1,2}$\orcid{0000-0002-7083-4049}}

\affil{$^1$Department of Astronomy, California Institute of Technology, Pasadena, CA 91125, USA}
\affil{$^2$Owens Valley Radio Observatory, California Institute of Technology, Big Pine, CA 93513, USA}
\affil{$^3$School of Earth and Space Exploration, Arizona State University, Tempe, AZ 85287, USA}

\affil{$^*$Author to whom any correspondence should be addressed.}

\email{rbyrne@caltech.edu}


\begin{abstract}
We identify a systematic distortion of the gain-vs.-frequency function of radio telescopes caused by digital flattening (``whitening'') of the signal's spectrum followed by re-quantization, a common pair of processes in the signal processing of modern telescopes. Wide-bandwidth telescopes often have a large variation of signal power over frequency. Flattening of the spectrum allows samples of the channelized signal to be represented in a small number of bits, allowing efficient downstream processing. However, we show that this produces subtle systematic error in the measured spectra. We explore this effect in data from the Owens Valley Radio Observatory's Long Wavelength Array (OVRO-LWA) and through detailed semi-analytic simulations. Although the effect can be small so that it has heretofore been unrecognized, we demonstrate that it produces distortion of the spectrum at a level that is problematic for some science, in particular 21 cm cosmology. Finally, we explore mitigation strategies, showing that the effect can be substantially reduced by careful choice of the gain distribution along the signal path or by incorporating dithering in the re-quantization step.
\end{abstract}

\section{Introduction} 

In radio telescopes with many antennas and wide-bandwidth receivers, it is often unavoidable that the signals received from antennas have a large spectral dynamic range; that is, their spectra are far from flat. This variation comes from the intrinsic frequency dependence of the sky signal, the antenna response, and filters in the signal path. However, when each signal is separated into many narrow-bandwidth frequency channels, each channel can have a small dynamic range. Therefore, its samples can be efficiently represented with only a few bits. To allow this, the power must be about the same in each channel, which requires that the spectrum be flattened or ``whitened''; this process is called equalization. The subsequent process of representing the samples in a small number of bits is called re-quantization (to distinguish it from the original quantization by the digitizer). 

\begin{figure}
    \centering
    \includegraphics[width=\linewidth]{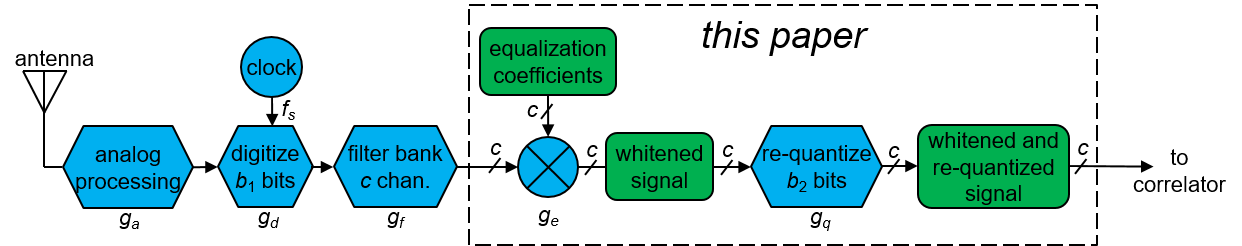}
    \caption{Block diagram of the typical per-signal processing architecture. Blue hexagons denote processing steps and green rectangles denote data and data products. This structure is applicable, with small variations, to many instruments. The signal from an antenna is amplified and filtered in analog circuitry, then sampled at rate $f_s$ and quantized to $b_1$ bits. It is then analyzed in a digital filter bank to produce $c$ frequency channels across the Nyquist band, each of bandwidth $f_s/(2c)$, after which all channels are processed in parallel. To flatten, or equalize, the spectrum, the time series of samples for each channel is multiplied by a constant coefficient that depends on the channel frequency, and finally each channel of the flattened signal is re-quantized so that it can be represented in a small number of bits $b_2$. The resulting time series are correlated against similar ones from all other antennas of the telescope. The dashed box shows the part of the processing that is analyzed in this paper. In the OVRO-LWA telescope, $b_1=10, c=4096$, and $b_2=4$.}
\label{fig:flowchart}
\end{figure}

A block diagram of equalization and re-quantization in the context of typical telescope processing is presented in Figure \ref{fig:flowchart}. The process analyzed here receives a time series of samples for each of $c$ frequency channels from a digital filter bank. The spectrum\footnote{Throughout this paper, we use the term ``spectrum'' as shorthand for the power spectral density function. Practical measurements via digital signal processing are given as the mean square of a series of samples in each of many discrete frequency channels.} is then flattened by multiplying each sample by a frequency-dependent but constant-time equalization function, and then re-quantized so that it can be represented in a small number of bits $b_2$. If the equalized sample value is too small or too large, it will underflow or overflow the $b_2$-bit number, respectively. To avoid this, the total gain from the antenna to the re-quantizer must be correct at each frequency. This includes the analog gain, which we denote $g_a$, and the digital gain $g_d g_f g_e g_q$. The latter is the product of the gains of all digital signal processing stages, where $g_d$ is the gain of the digitizer, $g_f$ is the gain of the filter bank, $g_e$ is the equalization gain, and $g_q$ is the re-quantization gain. We will see in subsequent sections that the distribution of the gain among its various factors is important.

\begin{figure}
    \centering
    \includegraphics[width=\linewidth]{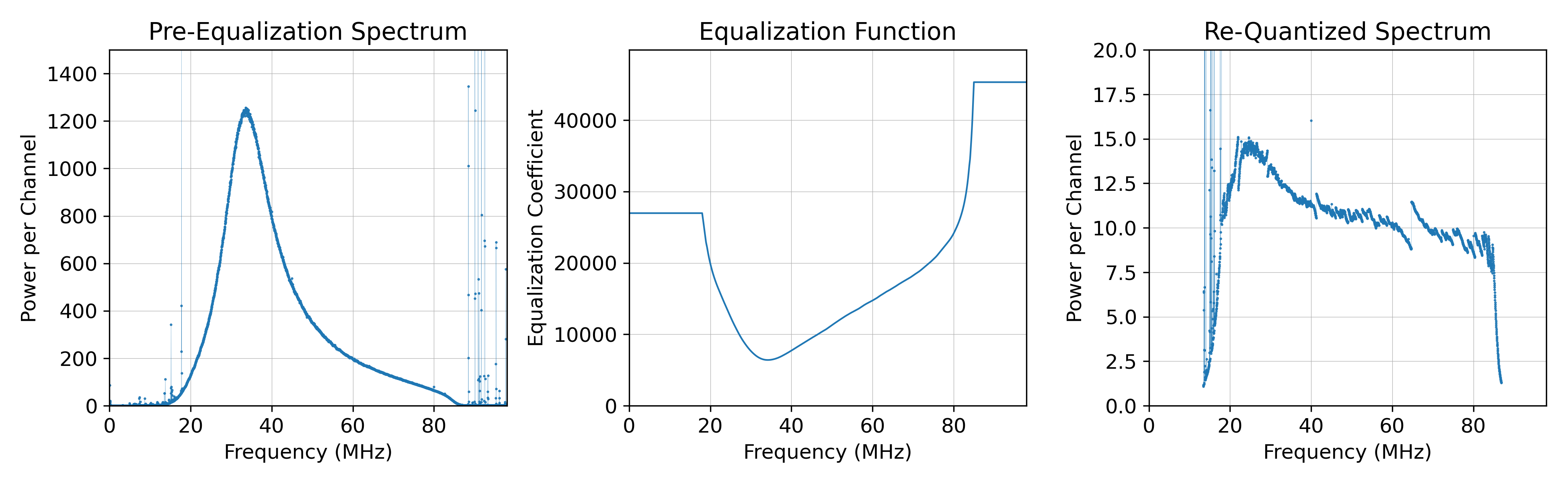}
    \caption{\textit{Left:} Spectrum of one signal from the OVRO-LWA, measured at the output of the filter bank before equalization and re-quantization. The signal has a large dynamic range across the full frequency range, owing to variations in the antenna sensitivity, filters in the analog signal chain, and intrinsic spectrum of the sky signal. Radio-frequency interference (RFI) causes strong, narrowband signals, visible as peaks in the spectra, that will be flagged and removed in later processing. \textit{Center:} Equalization function used for this signal. The equalization function designed to adjust the power per channel to a target value of 18, although signal variations cause the measured values to deviate from this target. Outside 18 to 83 MHz, the spectrum is falling rapidly; the equalization function is kept constant to avoid requiring a large dynamic range. \textit{Right:} Final spectrum, measured after equalization and re-quantization.}
\label{fig:eq_coeffs}
\end{figure}

The equalization and re-quantization of one signal of the Owens Valley Radio Observatory's Long Wavelength Array (OVRO-LWA) telescope \citep{eastwood2018, eastwood2019, anderson2019} is shown quantitatively in Figure \ref{fig:eq_coeffs}. This shows the measured spectrum (mean squared amplitude of the complex signal in each channel) at the filter bank output, the equalization function, and the measured spectrum after equalization and re-quantization\footnote{In radio astronomy, this is sometimes called an ``autocorrelation.''}. We use 4-bit re-quantization ($b_2=4$). 

\begin{figure}
    \centering
    \includegraphics[width=\linewidth]{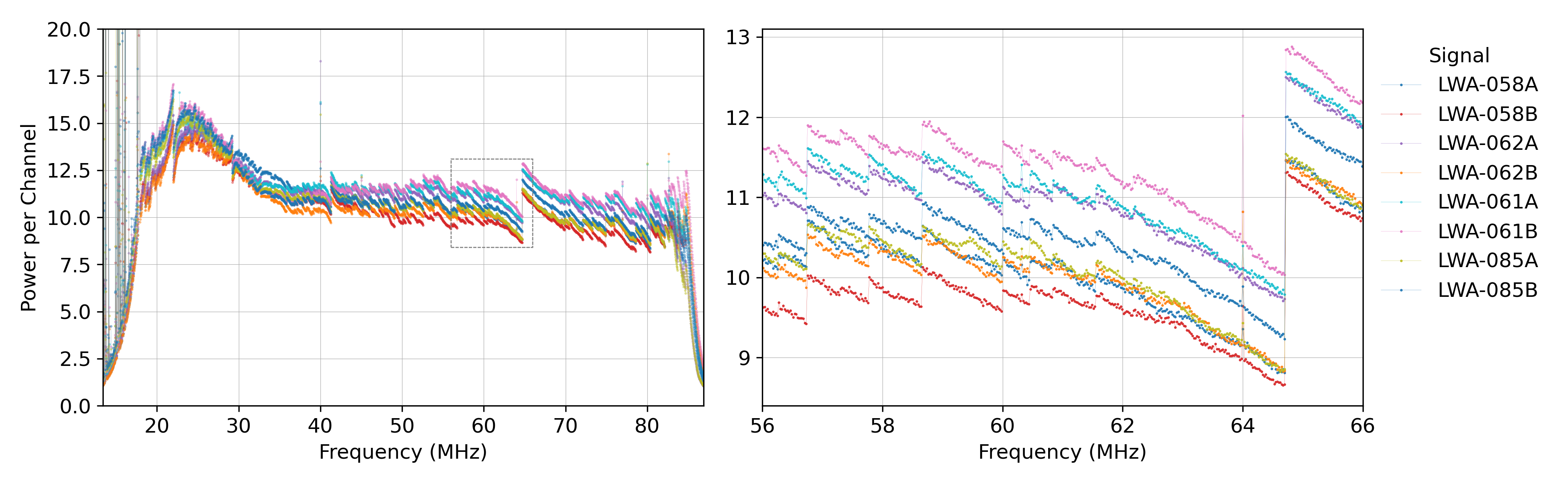}
    \caption{Measured spectra of 8 signals from the OVRO-LWA, after equalization and re-quantization. These are raw outputs from the correlator, prior to any calibration, using one 10 s time integration. We plot the spectra of individual signals (in the context of radio interferometry, these are sometimes called ``autocorrelations'' to distinguish from cross-correlated signals). The left panel shows the full spectrum, while the right panel zooms in. The spectra exhibit a discontinuous sawtooth pattern that emerges from digital equalization process, as explained in the text. All signals here use the equalization function plotted in the center panel of Figure \ref{fig:eq_coeffs}. The discontinuities occur at irregular intervals but are at the same frequencies for all signals. (Other signals in the OVRO-LWA that use different equalization functions have discontinuities at different frequencies.) Signals are labeled with an antenna number and polarization, denoted ``A'' (East-West aligned) or ``B'' (North-South aligned).}
\label{fig:autocorr_spec}
\end{figure}

In working with the OVRO-LWA telescope, we have noticed that subtle numerical effects in the digital equalization and re-quantization processes can, under some circumstances, introduce significant spectral distortion in the form of step changes in net gain at some frequencies, as seen in the right panel of Figure \ref{fig:eq_coeffs}. This is shown for additional signals and in more detail in Figure \ref{fig:autocorr_spec}. (Figures \ref{fig:eq_coeffs} and \ref{fig:autocorr_spec} depict spectra from individual signals only; \S\ref{s:impact_on_21cm} discusses the impact of this effect on interferometric visibilities.) The effect can be removed with some precision by astronomical calibration because the effect is systematic and repeatable. However, for science applications with especially stringent spectral-smoothness requirements, such as 21 cm cosmology, calibration may not deliver the precision to sufficiently mitigate the effect.

Neutral hydrogen cosmology uses the redshifted hyperfine emission line from HI (``21 cm emission'') to map large-scale structure throughout cosmic history \citep{Furlanetto2006, Morales2009, Pritchard2012, liu2020}. The signal is obscured by bright astrophysical foregrounds that are 4-5 orders of magnitude brighter. In principle, the cosmological signal and foregrounds can be separated based on the assumption that the foregrounds---synchrotron and free-free emission---are spectrally smooth. However, this requires extreme spectral fidelity to ensure that the instrument and analysis do not introduce spurious spectral structure. 

In recent years, the field has focused on two complementary approaches to mitigating spectral error. The first approach is to improve the accuracy of calibration through the development of detailed sky models (see, for example, \citealt{Carroll2016}, \citealt{Shimwell2017}, \citealt{eastwood2018}, \citealt{deGasperin2020}, \citealt{Lynch2021}, \citealt{Byrne2021}) and novel calibration techniques \citep{Liu2010a, Grobler2018, Dillon2018, Dillon2020, Byrne2021b, Ewall-Wice2022, Sims2022, Byrne2023}. The second approach is to develop instruments whose signal paths have gain-vs.-frequency functions that are inherently smooth \citep{thyagarajan2016, acedo2017, fagnoni2020, Cumner2023, ohara2024}. In practice, recent 21 cm experiments have used some combination of these approaches \citep{Patil2017, Barry2019, Abdurashidova2022, HERA2026}. This is because the best calibration techniques with the best sky models cannot achieve sufficient signal-to-noise ratio in the calibration observations unless the gain functions are sufficiently smooth before calibration.

This paper characterizes and explores mitigation strategies for one source of spurious structure in the gain-vs.-frequency response of each signal path of a radio telescope. These systematic spectral effects emerge from digital equalization and re-quantization in the digital signal path. We have carried out simulations and practical measurements of this effect using the OVRO-LWA and investigated potential solutions through changes in instrument design or operation. Since many telescopes use a similar processing architecture employing equalization and re-quantization (e.g., \citealt{parsons2008}; \citealt{kocz2014}; \citealt{prabu2015}; \citealt{vanderbyl2022}), our results should be widely applicable to current and future instruments, including the Square Kilometre Array (SKA; \citealt{weltman2020}), the Deep Synoptic Array (DSA; \citealt{hallinan2019}), and the next-generation Very Large Array (ngVLA; \citealt{selina2018}).

\section{How Does Digital Equalization Cause Spectral Distortion?}
\label{sec:overview}

In this analysis, we assume that all numbers are represented in fixed-point, twos-complement binary, as is typical in practice (including in the OVRO-LWA). We treat each number as an integer.

\begin{figure}
    \centering
    \hbox{
        \includegraphics[width=0.46\linewidth]{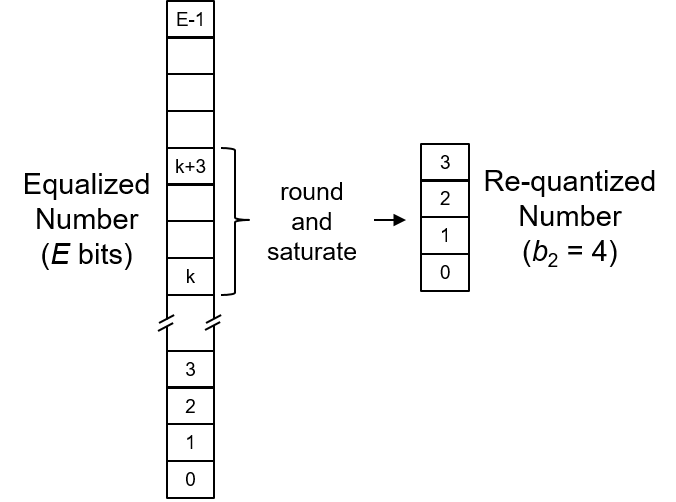}
        \includegraphics[width=0.54\linewidth]{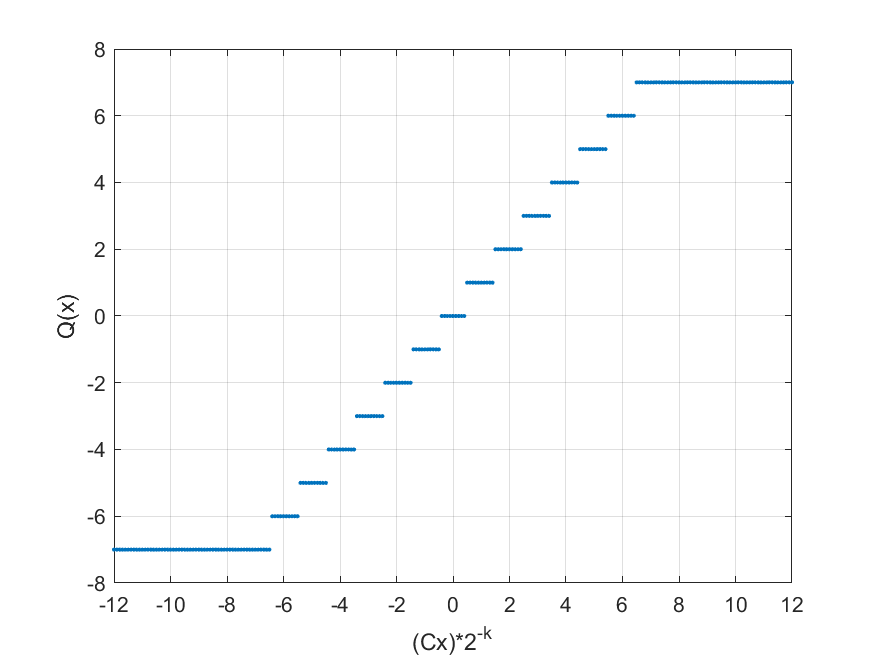}
    }
    \caption{{\it Left:}\ Re-quantization operation from an $E$-bit signed number to a $b_2$-bit signed number for $b_2=4$. Bit 0 of the re-quantized result corresponds to bit $k$ of the input. {\it Right:}\ Resulting transfer function for input value $x$ and equalization coefficient $C$.
    }
\label{fig:requantization}
\end{figure}

Let $x$ be an input sample to the equalization and re-quantization process from the filter bank for one frequency channel, and let $y$ be the corresponding output sample to the correlator (Figure \ref{fig:flowchart}). Then $y = Q(Cx)$, where $C$ is the equalization coefficient for that channel and $Q(\cdot)$ is the re-quanitzation function that maps its argument to one of the $2^{b_2}$ values that can be represented in $b_2$ bits. Many mappings are possible\footnote{
Mappings in which the output value is not encoded as a two-complement number can be more efficient and have been used in some telescopes, including the original VLA and various VLBI correlators \citep{TMS2001}. For example, $b_2=4$ allows the 16 values in [--8,+7] to be represented, but we use only 15 of them, omitting --8. This avoids bias. We could have achieved an unbiased mapping by omitting 0 instead, which would increase the dynamic range from $\pm7$ to $\pm8$, but then the output would not be interpretable as twos-complement binary. The only reason to constrain the output to twos-complement is to allow downstream processes like correlation to be implemented in commodity devices like CPUs and GPUs. Arithmetic for other number representations is efficient in FPGAs and ASICs, which were the favored devices for correlator implementations in the past \citep{daddario2016}.
}
but the one we use is illustrated in Figure \ref{fig:requantization}. The equalized value $Cx$ is a multi-bit number from which we select $b_2$ bits, starting at bit $k$. The least-significant bit of the result is rounded using an unbiased method such as round-to-even. If bits $k+b_2$ and $k+b_2-1$ are different, then the value overflows the output representation; in that case, the result is ``saturated'' at $+(2^{b_2-1}-1)$ or $-(2^{b_2-1}-1)$ depending on whether the sign of the input is positive or negative, respectively. If the input value is too small, the result underflows, so that the result is zero even though the input is non-zero. Since the input time series represents zero-mean random noise, the coarsely-quantized output is a statistically-accurate representation of the input provided that overflows and underflows occur only rarely. To achieve that, the variance of the equalized values $Cx$ must be in a narrow range. Ensuring that it is in that range for all frequency channels is the purpose of equalization. The variance depends not only on the values of the equalization function but also on the total gain from the antenna to the re-quantizer, including all the gain factors shown in Figure \ref{fig:flowchart}.

The gain of the equalization and re-quantization process is given by
\begin{equation}
	g_e g_q = \frac{\sigma_y^2}{ \sigma_x^2} = \frac{\sum_i P_i y_i^2}{\sum_i P_i x_i^2} = \frac{\sum_i P_i Q(Cx_i)^2}{ \sum_i P_i x_i^2}.
    \label{eq:eq_req}
\end{equation}
Here $x$ and $y$ have a finite number of possible values, indexed by $i$. $P_i$ is the probability that input $x$ has value $x_i$, and $y_i = Q(Cx_i)$. 

In the absence of overflows and underflows, $Q(Cx) \approx Cx\,2^{-k}$, where the approximation is due only to the rounding. In that approximation, the variance of the output is $\sigma_y^2 = (C2^{-k})^2\sigma_x^2$ , so the gain is $g_e g_q = (C2^{-k})^2$. This does not produce the spectral distortion in the re-quantized spectrum seen in Figure \ref{fig:autocorr_spec} provided that the input variance and the equalization function vary smoothly with frequency, which they normally do. The distortions occur when the input value $x$ is already coarsely quantized, prior to the desired re-quantization. Since $x$ is represented as a fixed-point number\footnote{
If instead the filter bank used floating point numbers with sufficiently large mantissas, the issue we are studying here could be largely
avoided. This is usually not practical in telescopes with many antennas and wide bandwidth because it would require excessive data rates and/or processing power. See further discussion in \S\ref{sec:mitigation}.
},
this happens when it is small. This means that it usually has only a small number of significant bits and a small number of likely values, regardless of the number of bits used to store it in the hardware. After multiplication by an equalization coefficient, the values (and variance) can be much larger, but they still have only a small number of likely values. That coarse quantization then interacts with the our desired re-quantization in such a way that a small change in equalization coefficient (from one frequency channel to the next) can produce a large change in output variance.

\begin{figure}
    \centering
    \includegraphics[width=\linewidth]{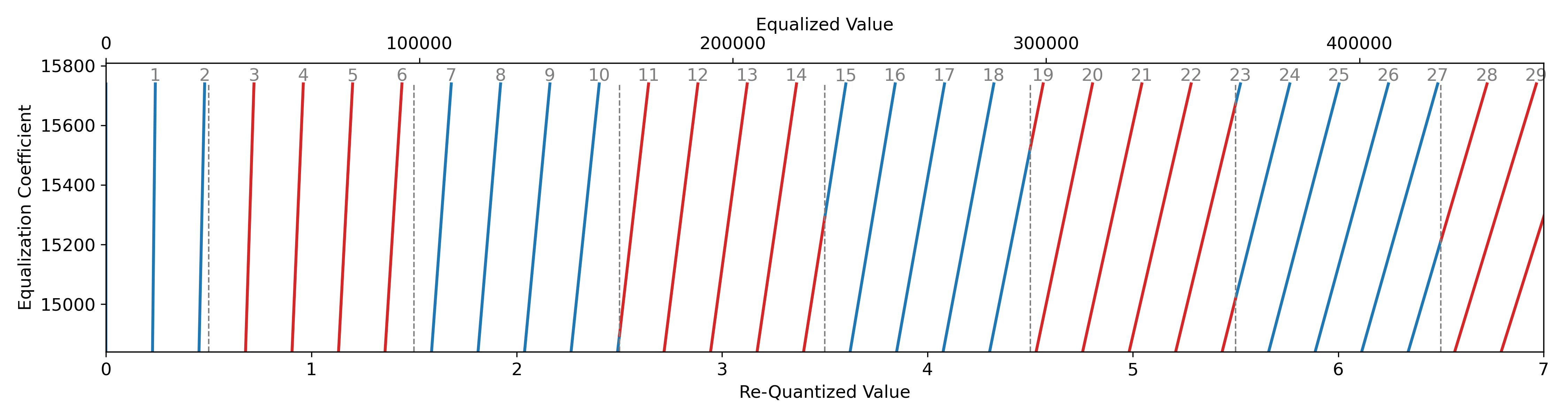}
    \caption{Depiction of the equalization and re-quantization process for a subset of possible values. Each diagonal line corresponds to the same pre-equalization value, from 0 to 29, as labeled. When these values are multiplied by an equalization coefficient (vertical axis) they yield an equalized value (top axis). All equalized values that fall between dashed vertical lines map to the same re-quantized value (bottom axis). Points along the blue lines map to even values and points along the red lines map to odd values. The plot corresponds to $k=16$ in Figure \ref{fig:requantization}.}
\label{fig:requant_drift}
\end{figure}

To see this, consider Figure \ref{fig:requant_drift}, which shows the $x$ to $y$ mapping for $b_2=4$ and $k=16$ (as used in the OVRO-LWA) over a portion of the range of pre-equalization values $x$ and equalization coefficients $C$. It shows how input values in [0, 15] map to output values in [0, 7] for coefficients in [14800, 15800]. (The mapping is an odd function, so it is symmetrical for negative $x$.) The pre-equalization values are small, necessitating large values of the equalization coefficients, so only 4 or 5 pre-equalization values can map to the same re-quantized value. Over some ranges of equalization coefficient, the mapping remains unchanged, with each pre-equalization value $x$ always corresponding to the same re-quantized value $y$, e.g.\ for $e=$ 15240 through 15510. Over such a range, the gain remains constant, even though the equalization coefficient is changing. At the boundaries of such ranges (e.g., near $e=15240$ in Figure \ref{fig:requant_drift}), the mapping changes; a small change in equalization coefficient then causes a large change in gain. In this way, the equalization and re-quantization process produces the spectral distortions seen in Figure \ref{fig:autocorr_spec}, even when both the input and the equalization function are spectrally smooth. 

Notice that the mapping illustrated in Figure \ref{fig:requant_drift} is the same for each equalization coefficient value, independent of the probability distribution of the input. Therefore, for the same equalization function the gain discontinuities always occur at the same frequencies. Since the equalization function is designed to produce constant output variance, the input probability distribution will always be approximately (but not exactly) the same near each gain discontinuity, so the magnitude of the discontinuity will stay approximately the same with time and across different signals in the array that use the same equalization function. Thus the effect is mostly time-invariant. Therefore, unlike quantization noise, it cannot be modeled as an additive random process whose effect on the spectrum diminishes with averaging time in accordance with the radiometer equation. 

\begin{figure}
    \centering
    \includegraphics[width=\linewidth]{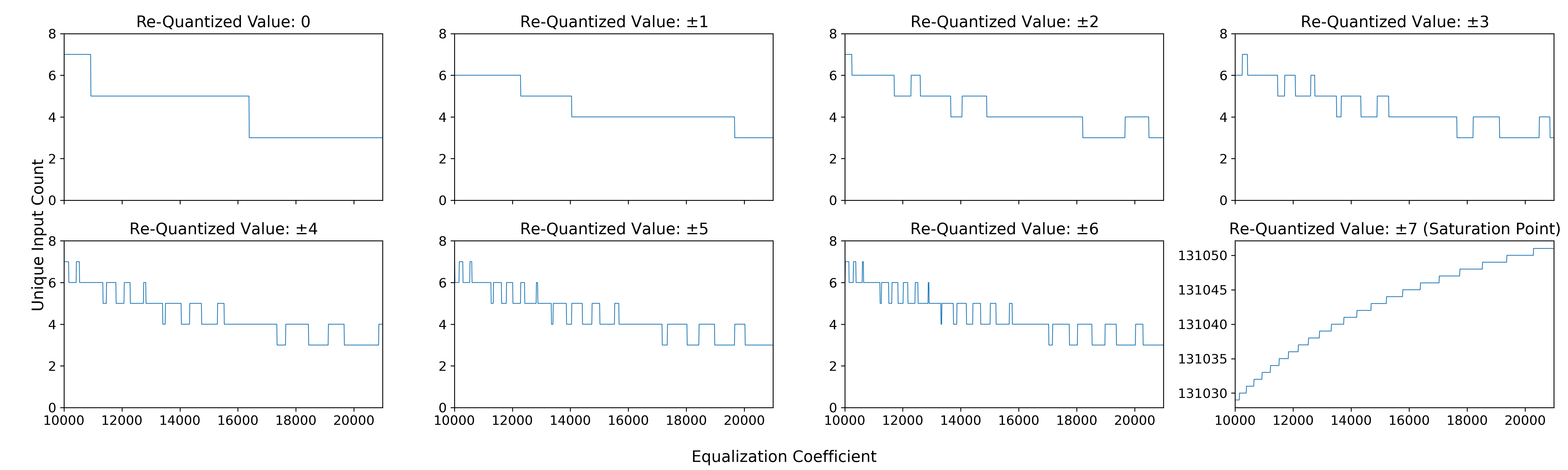}
    \caption{Histograms showing the number of unique input values that map to each 4-bit output value in re-quantization. The horizontal axes give the equalization coefficient for $k=16$. Changes in equalization coefficient cause discrete steps, as values cross the boundaries of the re-quantization bins. Note that, due to saturation, a large number of possible input values map to output values $\pm$7.}
\label{fig:requant_hist}
\end{figure}

Figure \ref{fig:requant_hist} elucidates the effect in a different way by presenting a histogram of the number of unique input values that map to each re-quantized value. As the equalization coefficient increases, fewer unique input values map to each re-quantized value in the range [-6, 6], and more values saturate. This process is not monotonic, and the discrete steps in these histograms occur at irregular intervals.

If the pre-equalization values were much larger (requiring much smaller equalization coefficients to get into the desired range), then many more than 4 or 5 of them would map to each re-quantized value, causing the step changes in gain to be far smaller. This is controlled by the distribution of gain along the signal path. Although the total gain required at each frequency is determined entirely by the spectrum at the antenna terminals, how that gain is distributed can be chosen by design. By placing more gain ahead of the equalizer ($g_a$, $g_d$, and $g_f$ in Figure \ref{fig:flowchart}) and less after that point ($g_e$, $g_q$), the size of the discontinuities can in principle be made as small as desired. In practice, however, this is subject to constraints. If the analog gain $g_a$ is too large, the digitizer can saturate frequently. After digitization, if the gain is too large at any stage, the resulting numbers can overflow their finite-bit-width representations.

In the OVRO-LWA, radio frequency interference (RFI) from human-made sources is often very strong. While much of it is outside the frequency range of interest (roughly 15 to 85 MHz) and is suppressed by analog filters, a significant amount reaches the digitizer and this limits the maximum analog gain $g_a$. It also affects the design of the filter bank, which has a polyphase architecture (PFB) \citep{Crochiere1983} that uses 18-bit fixed-point number representations internally and at its output. Bit growth during the multiple processing stages of the PFB can easily cause overflow if any frequency channel has too much power. Although channels containing RFI can be ignored, overflows can contaminate many other channels. Overflows are avoided by scaling down (right-shifting) the results after each stage, but this prevents the PFB gain $g_f$ from being large. It is these constraints that cause the pre-equalization sample variances to be small, leading to the spectral distortions at issue here. In other instruments with similar processing architectures, the distortions may be weaker, depending on the distribution of gain along the signal path\footnote{
The distortions were also much weaker during early commissioning of the OVRO-LWA, delaying recognition of their importance. The analog gain was then higher and equalization function smaller, but eventually the telescope experienced very strong in-band RFI in daytime due to ionospheric reflection of distant transmitters. To mitigate this, the analog gain is now decreased and equalization function increased during daytime. 
}.



\section{Simulating Digital Equalization and Re-quantization} \label{sec:simulation}

We present a semi-analytic simulation method that captures equalization and re-quantization in the digital signal path presented in Figure \ref{fig:flowchart}. Adjusting the simulation parameters allows us to investigate the impact of various mitigation strategies.

Early versions of our simulation took a Monte Carlo approach, where we used random input values and evaluated the result across many trials. To eliminate sample variance without requiring an infeasibly large number of trials, we moved to a semi-analytic model in which we propagate a probability distribution through the simulated signal chain. The results correspond to the limit of infinite trials, or arbitrary long accumulation times.

\subsection{Representing the Filter Bank Output}\label{s:quantized_signal_sim}

The input to the simulation is the filter bank output signal (see Figure \ref{fig:flowchart}). For each frequency channel, we assume that the time series consists of samples from a zero-mean Gaussian random process represented as twos-complement, fixed-point binary numbers. We model each sample of the underlying process with a probability density function given by
\begin{equation}
    P(x) = \frac{1}{\sqrt{2 \pi} \sigma} e^{-x^2/(2 \sigma^2)},
\label{eq:gaussian_distribution}
\end{equation}
where $\sigma$ is the standard deviation. This is an approximation; the true probability density function will deviate from a Gaussian due to upstream nonlinear quantization effects not considered in this analysis. 

To represent $x$ as two-complement binary with $b$ bits, let $i$ be an unbiased rounding of $x$ to the nearest integer, saturated at $i_\text{min}=-2^{b-1}$ and $i_\text{max}=+2^{b-1}-1$. The discrete random variable $i$ then has a probability distribution given by
\begin{equation}
    P_i = \begin{dcases} 
    \frac{1}{\sqrt{2 \pi} \sigma} \int_{i-\frac{1}{2}}^{i+ \frac{1}{2}} e^{-x^2/(2\sigma^2)}\,dx = \frac{1}{2} \left[ \erf\left(\frac{i+\frac{1}{2}}{\sqrt 2\sigma}\right) - \erf\left(\frac{i-\frac{1}{2}}{\sqrt 2\sigma}\right)\right], & \mbox{ if } i_\text{min} < i < i_\text{max} \\
    \frac{1}{\sqrt{2 \pi} \sigma} \int_{-\infty}^{i+\frac{1}{2}}e^{-x^2/(2\sigma^2)}\,dx = \frac{1}{2} \erf\left(\frac{i+\frac{1}{2} }{\sqrt 2\sigma}\right), & \mbox{ if } i = i_\text{min} \\
    \frac{1}{\sqrt{2 \pi} \sigma} \int_{i-\frac{1}{2}}^\infty e^{-x^2/(2\sigma^2)}\,dx = \frac{1}{2} \erf\left(\frac{i-\frac{1}{2}}{\sqrt 2\sigma}\right), & \mbox{ if } i = i_\text{max} \\
    \end{dcases},
\label{eq:quantized_pdf}
\end{equation}
where $\erf(\cdot)$ denotes the error function.

The average power in a random process is given by its variance
\begin{equation}
    p = \text{var}(i) = 2\sum_{i=i_\text{min}}^{i_\text{max}} P_i\,|i|^2,
    \label{eq:variance}
\end{equation}
and doing this for each channel at frequency $f_n$ gives its spectrum $p(f_n)$. Each channel of the filter bank output is actually a complex-valued random sequence, but the distribution in Equation \ref{eq:quantized_pdf} applies to a real random variable. We assume that its real and imaginary parts are independent and identically distributed; this produces the factor of 2 in Equation \ref{eq:variance}.

\subsection{Equalization}

Equalization, or whitening, multiplies each frequency channel in the signal path by a real-valued constant. We simulate this process by propagating the discrete probability distribution (Equation \ref{eq:quantized_pdf}) through this multiplication, giving spectrum
\begin{equation}
    p_\text{eq}(f_n) = C(f_n)^2\,p(f_n),
\end{equation}
where $C(f_n)$ is the equalization coefficient.  

$C(f_n)$ is represented in fixed-point binary with $b_C$ bits. Recalling that $i$ is represented with $b$ bits (real and imaginary parts separately), to represent the product $C(f_n)i$ for each sample without overflow requires at least $b+b_C$ bits. This is done in the OVRO-LWA hardware and also in our simulation, using $b_C=16$. 

\begin{figure}
    \centering
    \includegraphics[width=\linewidth]{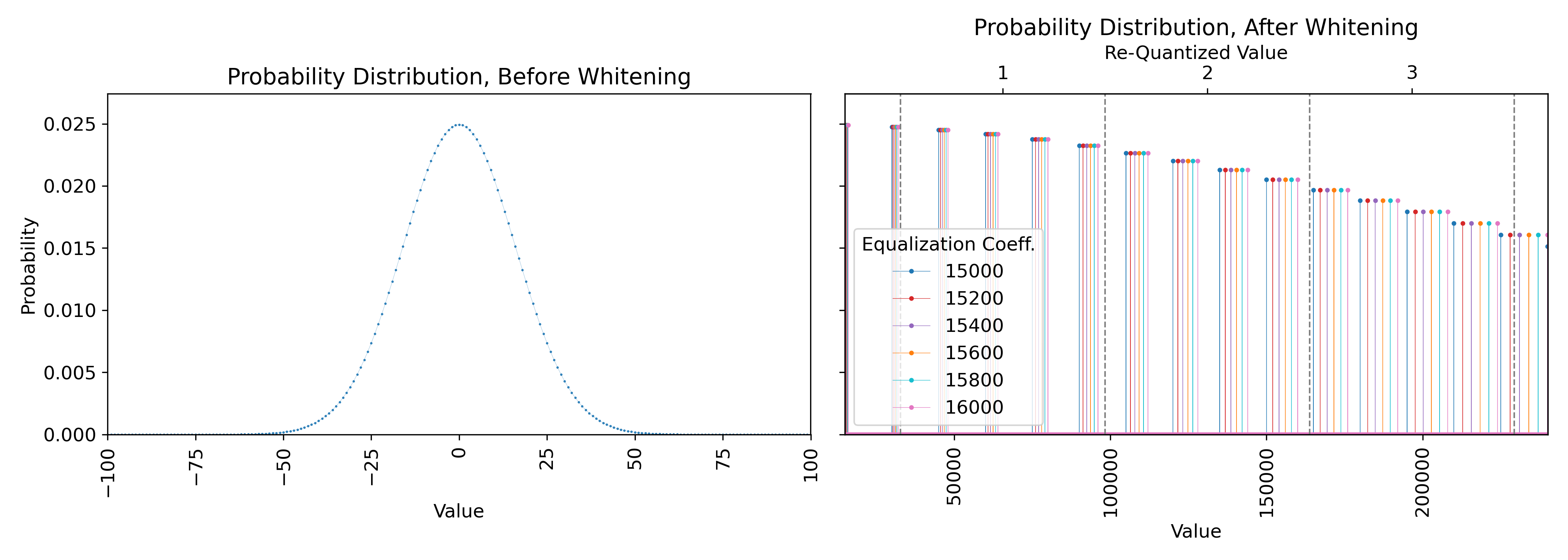}
    \caption{Simulated probability distribution before equalization (left) and after equalization (right). The former is for a discrete-valued Gaussian with a standard deviation of $\sigma=16$, as calculated by Equation \ref{eq:quantized_pdf}, representing either the real or imaginary part of a simulated filter bank output. At each frequency, it is multiplied by constant equalization coefficient in the equalization step. The equalization function is frequency-dependent, but here we consider just one frequency channel. The probability distribution of the equalized signal (right) has a comb-like structure: only multiples of the equalization coefficient have non-zero probability. Here we plot six distinct equalization coefficients and show just a small range of values on the horizontal axis in order to highlight the probability distribution's discrete structure. 
    The vertical dashed lines denote boundaries of the re-quantization bins for the specific case of $k=16$ and $b_2=4$ (see Figure \ref{fig:requantization}); the top axis labels the re-quantized values. See section \ref{sec:requantization} for discussion.
    }
\label{fig:pdf_whitening}
\end{figure}

The probability distribution of the signal before and after equalization is plotted in Figure \ref{fig:pdf_whitening} for $\sigma = 16$, $b=18$, and some large values of $C(f_n)$. Whereas the input is weak ($\sigma \ll 2^{b-1}$), equalization requires large values of the coefficients $C(f_n)$. This means that many values of the product have zero probability.

\subsection{Re-quantization} \label{sec:requantization}

\begin{figure}
    \centering
    \includegraphics[width=\linewidth]{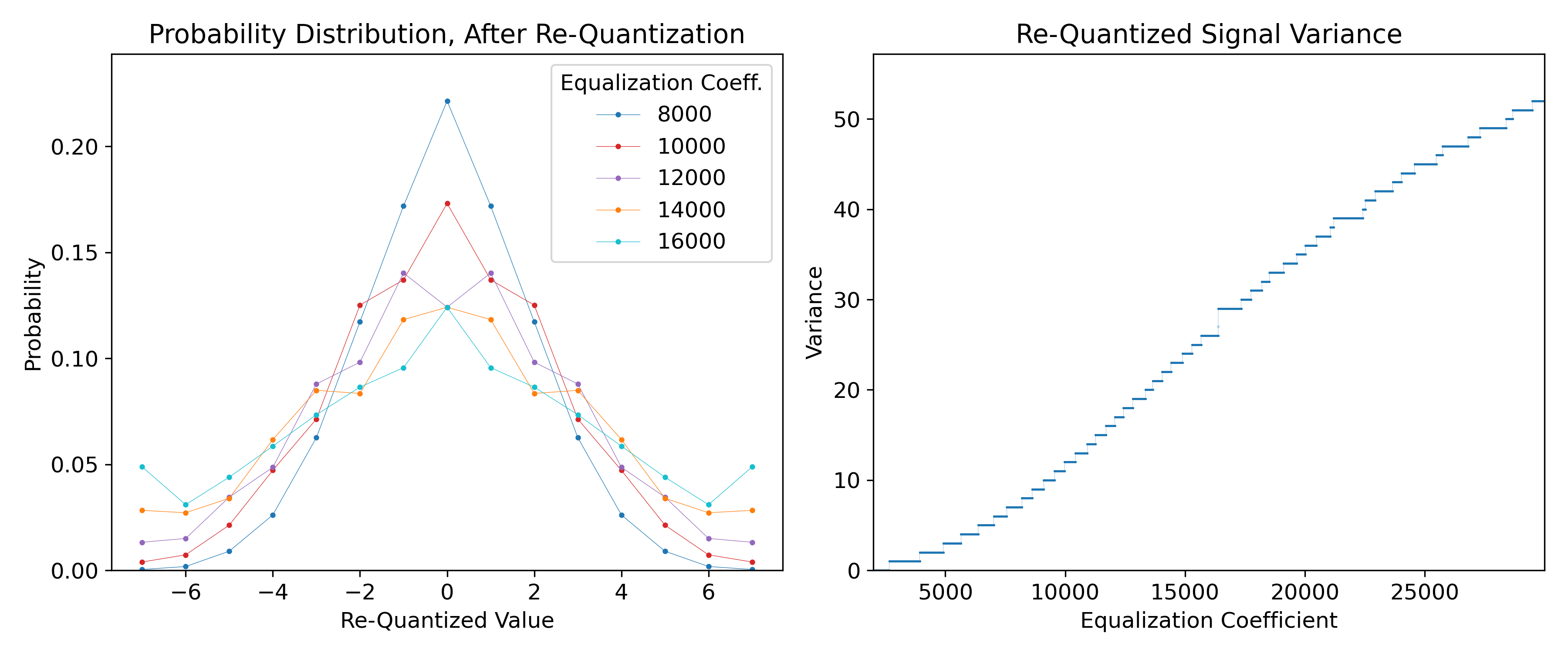}
    \caption{\textit{Left:} Probability distribution of the re-quantized values for the same parameters as in Figure \ref{fig:pdf_whitening} ($\sigma=16$, $k=16$, $b_2=4$), plotted for 5 distinct equalization coefficients. For a constant input signal, larger equalization coefficients broaden the distribution. The re-quantized values saturate at $\pm 7$. \textit{Right:} From Equation \ref{eq:variance}, we calculate the variance of the re-quantized signal. For a constant input signal, the variance increases with equalization coefficient, but the increase is stepwise and occurs at irregular intervals.}
\label{fig:pdf_requantized}
\end{figure}

Re-quantization reduces the number of bits in the real and imaginary parts of the equalized values by mapping each equalized sample value to one of the possible re-quantized values (see Figure \ref{fig:requantization}).

In simulation, we propagate the equalized probability distribution shown in Figure \ref{fig:pdf_whitening} through the re-quantization process. This requires summing the probabilities of all equalized sample values that map to the same re-quantized value. This is illustrated in the right panel of Figure \ref{fig:pdf_whitening} for the case of $k=16$ and $b_2=4$, where the vertical dashed lines denote boundaries of the re-quantization bins.

The left panel of Figure \ref{fig:pdf_requantized} shows the resulting probability distribution for five distinct values of the equalization coefficient. Since the variance of the input signal is held constant, with $\sigma=16$, increasing the equalization coefficient broadens the distribution. From Equation \ref{eq:variance}, we can relate this probability distribution to an expected variance, plotted in the right panel of Figure \ref{fig:pdf_requantized}. The variance shows irregular, stepwise variation as the equalization coefficient varies. 

\subsection{Simulating the OVRO-LWA}
\label{sec:simulation_results}

With the simulation described here, we can now simulate the OVRO-LWA signal through the equalization and re-quantization steps. 

\begin{figure}
    \centering
    \includegraphics[width=\linewidth]{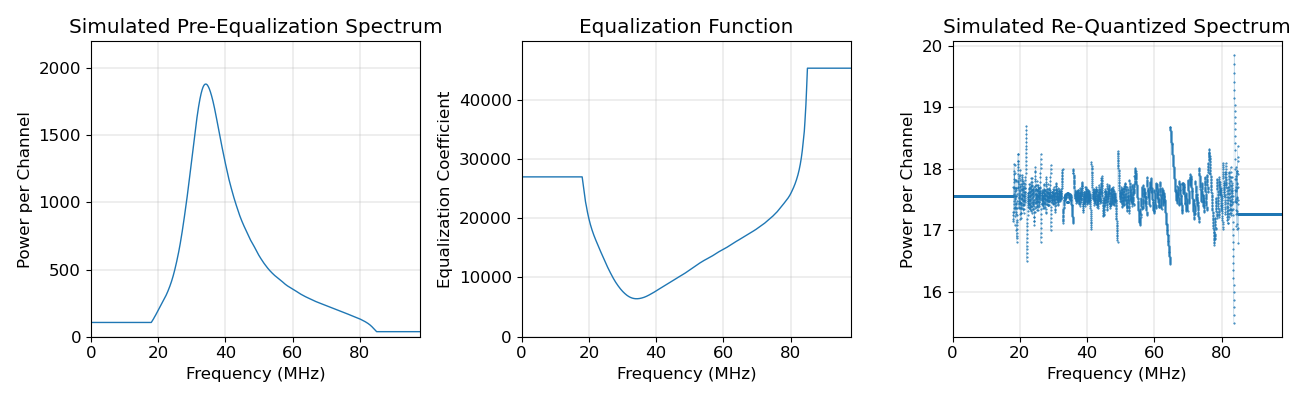}
    \caption{Simulation of the OVRO-LWA signal. \textit{Left:} The simulation input, corresponding to the filter bank output (see Figure \ref{fig:flowchart}). Our simulation assumes that the filter bank output is a Gaussian random signal with a frequency-dependent variance, plotted here. \textit{Center:} We use the same equalization function as in Figure \ref{fig:eq_coeffs}. We assume that the equalization function is perfectly matched to the data, such that the signal in the left panel is proportional to the reciprocal of the equalization function. \textit{Right:} The simulated spectrum of the equalized and re-quantized signal. The signal is propagated through a noiseless, semi-analytic simulation of the equalization and re-quantization steps. Although the simulation input and the equalization function vary smoothly with frequency, the re-quantized spectrum shows sharp discontinuities. These discontinuities are a systematic effect that are identical for all time steps, so they are not mitigated by accumulating the data in time.}
\label{fig:lwa_sim}
\end{figure}

We use the same equalization function plotted in Figure \ref{fig:eq_coeffs}. The equalization function was derived empirically by measuring the spectrum of the filter bank output signal and targeting a standard deviation of $\sigma_q=3.0$ for the re-quantized 4-bit values. In simulation, we assume that the filter bank output is a Gaussian random signal with zero mean and a frequency-dependent variance. We set that variance so that the target re-quantized variance will be achieved exactly:
\begin{equation}
    \sigma(f_n) = (\sigma_q \times 2^k) / C(f_n),
\label{eq:stddev_vs_eq_coeff}
\end{equation}
where $C(f_n)$ are the equalization coefficients and $k$ is the least significant bit retained in re-quantization (see Figure \ref{fig:requantization}). The current OVRO-LWA implementation uses $k=16$ and we use that value in these simulations. The simulated filter bank output is plotted in the left panel of Figure \ref{fig:lwa_sim}.

The right panel of Figure \ref{fig:lwa_sim} shows the result of propagating this signal through the equalization and re-quantization steps. The simulated spectrum shows sharp discontinuities in frequency at irregular intervals. This agrees qualitatively with what we see in the OVRO-LWA data (Figure \ref{fig:autocorr_spec}).

\section{Impact on Analyses for 21 cm Cosmology}
\label{s:impact_on_21cm}

The simulations in \S\ref{sec:simulation} demonstrate that equalization and re-quantization can produce spectral discontinuities. In this section, we explore the impact of these discontinuities on 21 cm cosmology.

21 cm cosmology analyses fall into two categories: interferometric analyses aim to measure spatial fluctuations in the cosmological HI signal, while global analyses measure the sky-averaged signal. The principal challenge of each of these analyses is the same: to effectively separate the faint cosmological signal from the bright intervening foregrounds. In both cases, this requires a dynamic range of one part in $10^4-10^5$, and any source of spectral contamination that prevents achieving that dynamic range is at a level that swamps the 21 cm signal.

\begin{figure}
    \centering
    \includegraphics[width=\linewidth]{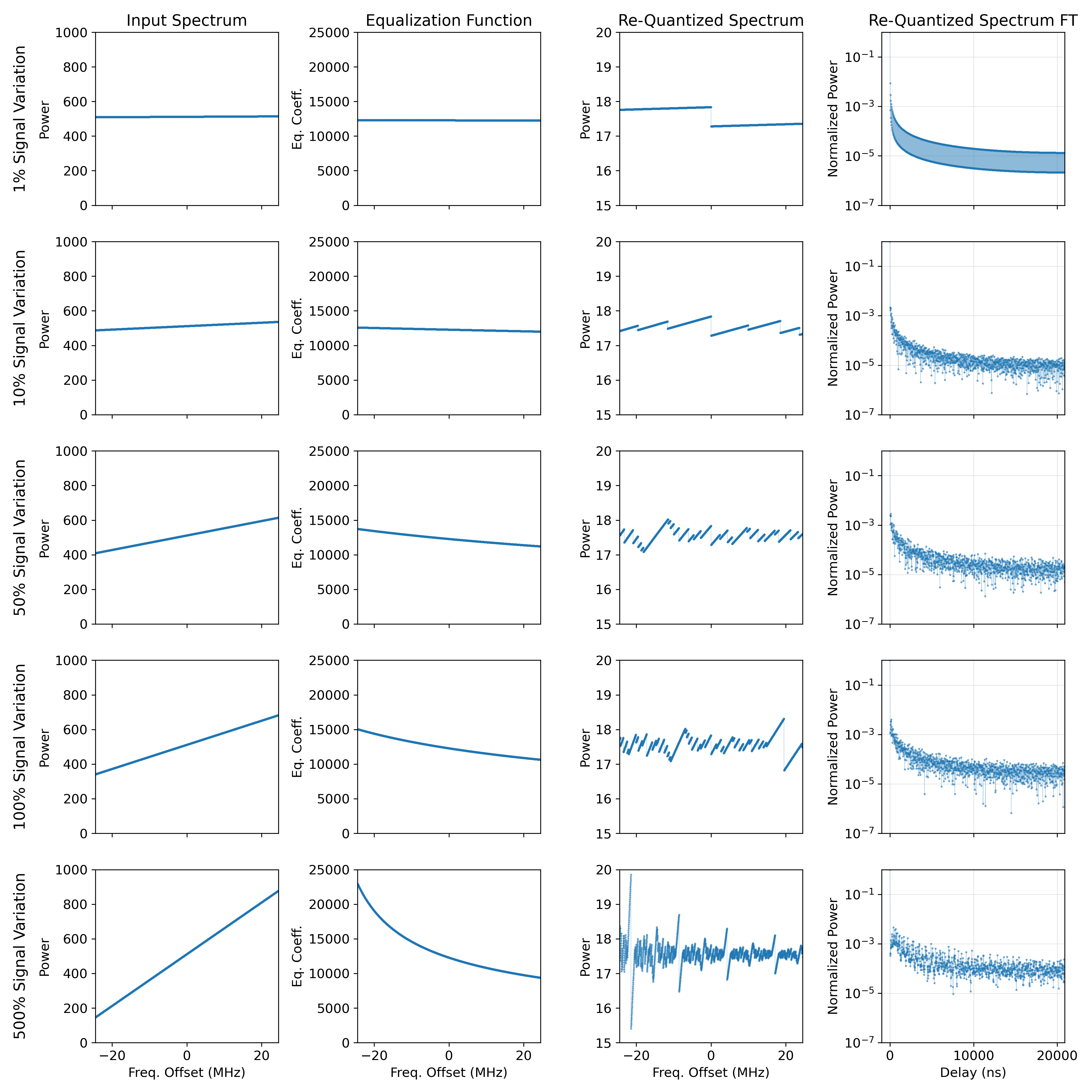}
    \caption{Simulated signals with linear variation of power with frequency. In the left column, we plot the variance of the simulation input, which is the output of the filter bank in the signal path (see Figure \ref{fig:flowchart}). We assume that the input is a complex Gaussian random signal that is quantized to 18 bits in its real and imaginary parts. From top to bottom, the four rows have slopes of 1\%, 10\%, 50\%, 100\%, and 500\% across 48.97 MHz. We constrain the mean power to be 512 in all cases. The second column plots the equalization functions used to flatten the signal. The third column shows the simulated spectrum of the 4-bit re-quantized signal, and the right column plots its Fourier transform, peak-normalized to 1. For nonzero delay, nonzero values in the Fourier transformed spectrum indicate spectral contamination that could affect the 21 cm cosmological measurement. Values greater than $10^{-5}$ are expected to swamp the faint cosmological signal.}
\label{fig:constant_slope_sims}
\end{figure}

In order to draw general conclusions about the impact of equalization and re-quantization, we simulate fiducial signals that vary linearly with frequency. We simulate a set of four signals with dynamic ranges varying from 1\% to 500\%, plotted in the left column of Figure \ref{fig:constant_slope_sims}. As in \S\ref{sec:simulation}, we represent the simulation input, corresponding to the filter bank output, as a Gaussian random complex signal that is quantized to 18 bits in each the real and imaginary parts. All four simulations have an average power of 512. We use a total frequency range of 48.97 MHz, corresponding to 2048 frequency channels with a channel width of 24 kHz (equal to the OVRO-LWA's frequency resolution). The simulation results are invariant to the overall frequency, so we present the results in with respect to the offset, in MHz, from the central frequency. 

We simulate an equalization function that is matched to the data, such that the frequency variation is completely eliminated prior to re-quantization. We once again use a target standard deviation of 3.0 (target power of 18.0) and $k=16$, and the relationship between the data standard deviation and the equalization function is given by Equation \ref{eq:stddev_vs_eq_coeff}. The equalization functions are plotted in the second column of Figure \ref{fig:constant_slope_sims}. To simplify interpretation of the results, and as a departure from more physically-realistic simulations of the OVRO-LWA signal path, we represent the equalization coefficients as floating point values, and the equalization procedure multiplies the 18-bit data at floating point precision. This means that the results are invariant to any quantization effects from the quantization of the equalization function.

Using the simulation method described above in \S\ref{sec:simulation}, we derive the spectrum of the 4-bit re-quantized signal, plotted in the third column of Figure \ref{fig:constant_slope_sims}.

The right column of Figure \ref{fig:constant_slope_sims} presents the Fourier transform of the spectrum across frequency, transforming it into delay space in units of time. We normalize the signal such that the peak value is 1. Away from a delay of zero, nonzero values indicate spectral contamination due to equalization and re-quantization.

These peak-normalized delay spectra provide a measure of the dynamic range, and we find that it does not reach $10^{-5}$ for all relevant delays in any of those simulations. This is true even for the case that the equalization function compensates for a signal variation of just 1\% across nearly 50 MHz. This illustrates the fact that the effect causes a small change in equalization coefficient to produce a much larger change in the re-quantized power, as explained in \S\ref{sec:overview}, so that it can be important even when the pre-equalization power varies very slowly with frequency.

Of course, the results presented in Figure \ref{fig:constant_slope_sims} depend on the simulation parameters used. The total power in the input signal is the same in all cases, making the RMS signal small at some or all frequencies. Increasing the signal amplitude would reduce spectral contamination, as discussed in \S\ref{sec:overview} and explored below in \S\ref{sec:increase_gain}.

For an interferometric array like the OVRO-LWA, the main observable is a visibility formed by cross-correlating two signals. We have shown that the equalization and re-quantization process causes distortion of the gain-vs.-frequency function of each signal, which leads to distortion of its spectrum. Each of the spectra plotted in Figure \ref{fig:constant_slope_sims} corresponds to one signal. In the context of an interferometric array, these are the same as correlating each frequency channel of a signal with itself. The two signals that contribute to a cross-correlated visibility may have the same or different equalization functions. The spectral distortion from the equalization and re-quantization process is likely to be worse when the equalization functions for the two signals are the same because, in that case, the distortions are aligned in frequency and tend to reinforce. In the OVRO-LWA, the required equalization functions are similar among signals, so that 7 different functions are sufficient for all 704 signals. It is unlikely that introducing intentional small variations of the equalization function among signals can provide sufficient improvement to appreciably mitigate the impact of the effect on interferometric 21 cm analysis.


In this treatment, we have not considered the role of frequency-dependent calibration. Calibration could mitigate spectral contamination, although it is unlikely that it can completely eliminate the effect. For example, \citealt{Barry2016} demonstrates that typical per-frequency sky-based calibration would require an infeasibly precise sky model in order to deliver the dynamic range required for 21 cm cosmology. Notably, the frequency discontinuities shown here preclude any calibration approach that assumes slow or continuous gain variation across frequency.

\section{Mitigation Strategies}
\label{sec:mitigation}

This section explores strategies for reducing systematic spectral distortion from equalization and re-quantization. These strategies include avoidance approaches that eliminate the need for equalization and mitigation approaches that reduce the amplitude of the systematic. 

\subsection{Analog Whitening}

The systematic frequency structure investigated in this paper occurs during digital equalization and re-quantization. One solution would is to whiten the signal prior to digitization. If the analog whitening filter appropriately matches the signals, this removes the need for digital equalization.

Retrofitting an instrument with analog whitening filters for each signal path (for the OVRO-LWA, this is 704 signal paths corresponding to two polarizations from each of 352 antennas) is generally infeasible. Even if it were possible, the natural spectrum is complicated and differs among antennas. While digital equalization parameters can be easily updated, analog whitening is less flexible. This means that it cannot be easily adapted to changing instrumental and environmental conditions, such as gain variations based on temperature or RFI conditions. An analog whitening filter could make the spectrum somewhat whiter but may not achieve the precision needed to avoid additional equalization after digitization. Furthermore, installing analog whitening filters would be prohibitively expensive and require significant downtime for the instrument.

\subsection{Floating Point Implementation}

If the filter bank and equalization were implemented with floating point computation, the issue we are studying here would be largely avoided. This could eliminate the chance of overflows, even in the presence of strong RFI at some frequencies, without requiring a reduction in the filter bank gain $g_f$ (see discussion in \S \ref{sec:overview} and below in \S \ref{sec:increase_gain}). Floating point computation poses challenges in wide-bandwidth telescopes with many antennas because of the additional processing burden and higher power consumption, so most telescopes (including ours) have used fixed-point implementations. However, a recent upgrade to the MWA telescope (\citealt{morrison2023}; S.\ Tingay et al.\ in prep.) uses a floating point FFT in its fine channelization step, implemented in GPUs rather than FPGAs.

\subsection{Piecewise-Constant Equalization}

If equalization were not needed, the spectral distortions we are studying would not occur. For many wide-bandwidth telescopes, equalization cannot be avoided over the full frequency range, but it might be avoidable over limited-bandwidth segments. This leads to the concept of piecewise-constant equalization, where the equalization function is constant over each segment but changes abruptly at segment boundaries. Then the overall gain remains constant within each segment, avoiding the spectral distortions, in exchange for large discontinuities in gain between segments.

A piecewise-constant equalization function means that 21 cm cosmology analyses must be performed separately for data within each interval, as these analyses cannot span the spectral discontinuities (for more detail about these analyses, see, for example, \citealt{liu2020} and \citealt{morales2012}). This potentially limits the sensitivity of the analysis, as it restricts the amount of data that can contribute to each measurement. Each interval corresponds to a redshift range, and these ranges are defined by the equalization function and cannot be modified after data capture.

\begin{figure}
    \centering
    \includegraphics[width=0.5\linewidth]{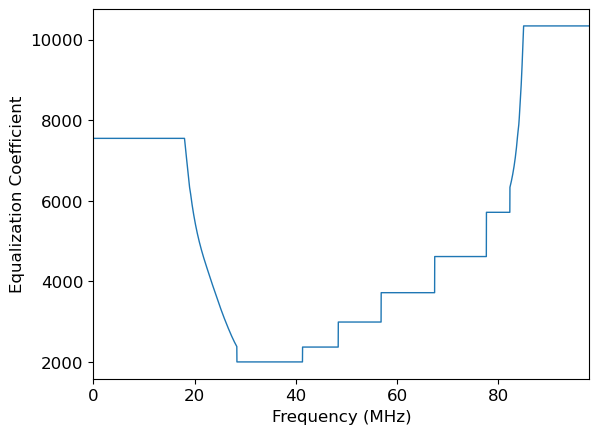}
    \caption{Example of a piecewise-constant equalization function used for the OVRO-LWA (compare to the previously implemented equalization function, plotted in Figure \ref{fig:eq_coeffs}). The equalization function is piecewise-constant across the frequencies of interest for the 21 cm cosmology analysis. The analysis is performed separately within each constant regions. Within those intervals, the coefficients need not be identical between antennas but must be held constant across frequency. The sizes of the frequency intervals were chosen to be as large as possible, in order to maximize the cosmological signal sensitivity, while avoiding excessive overflow and underflow of the 4-bit signal.}
\label{fig:eq_coeffs_piecewise}
\end{figure}

Despite its drawbacks, this approach is the simplest to implement. We have therefore decided to use piecewise-constant equalization functions for the OVRO-LWA, beginning in November 2025. An example is plotted in Figure \ref{fig:eq_coeffs_piecewise}.

\subsection{Increased Number of Bits in Re-quantization}

The number of re-quantization bits $b_2$ is limited by the capacity of the data link to the correlator (Figure 1), which requires transposing from all frequencies with each signal to all signals with each frequency and is usually implemented using a packet-switched network. In this section, we explore in simulation the extent to which increasing this number of bits can mitigate systematic spectral effects. Our simulations indicate that the improvement is minimal, showing that the effect explored in this paper is not a result of insufficient bits after re-quantization. Rather, as discussed in \S\ref{sec:overview} and explored below in \S\ref{sec:increase_gain}, it emerges because of coarse quantization of the signal \textit{prior} to re-quantization.

\begin{figure}
    \centering
    \includegraphics[width=\linewidth]{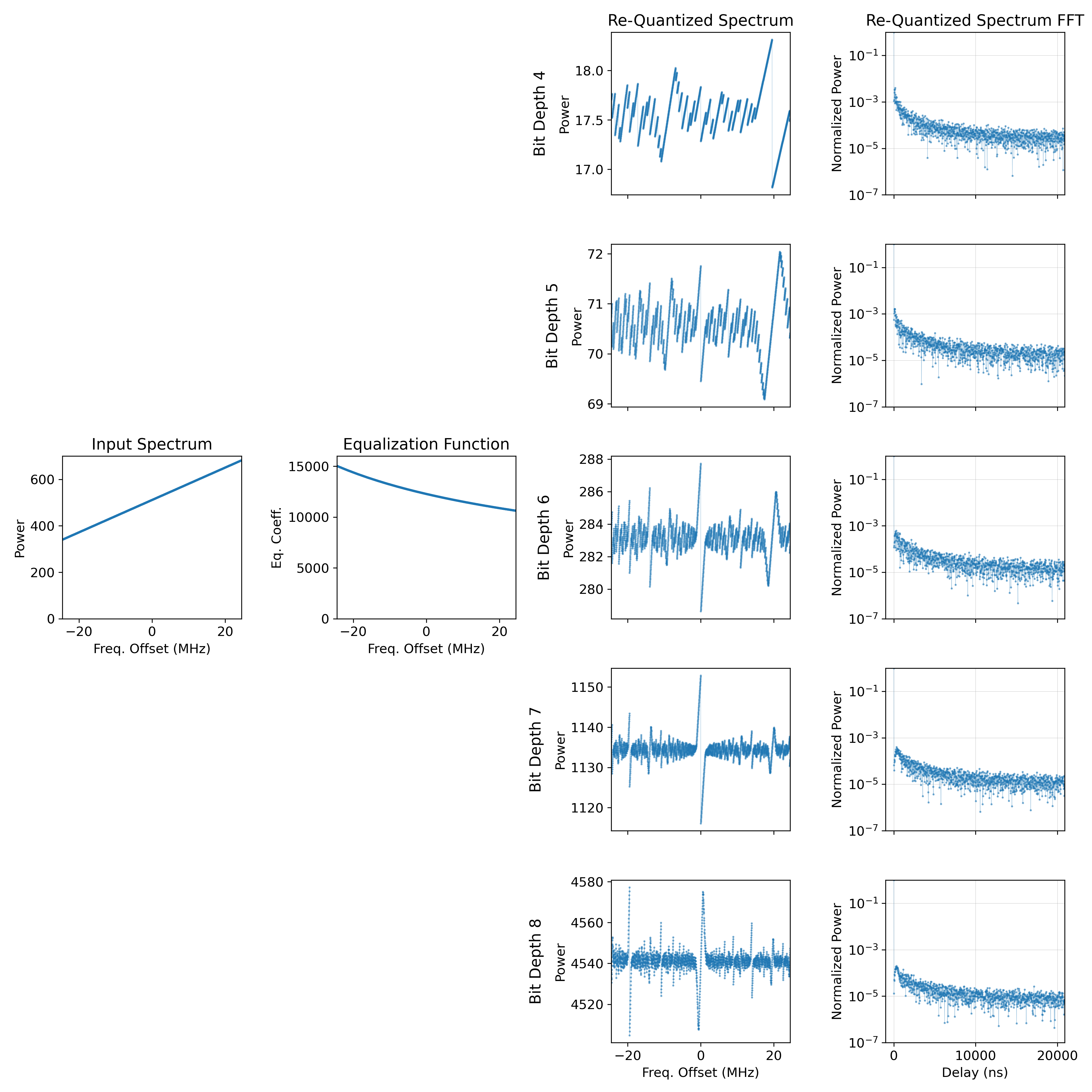}
    \caption{Simulated signals with varying numbers of output bits. Here the input signal and equalization function are held constant across all trials and are equivalent to the fourth row in Figure \ref{fig:constant_slope_sims}. The second from the right column shows the spectra for 4, 5, 6, 7, and 8 bits and the rightmost column shows the normalized Fourier transform of the spectrum. The average power increases by approximately a factor of 4 with each additional bit. Increasing the number of re-quantization bits modestly reduces spectral contamination.}
\label{fig:bit_depth_var_sims}
\end{figure}

The simulation results are plotted in Figure \ref{fig:bit_depth_var_sims}. Each trial uses the same input signal, consisting of a monotonically increasing spectrum, and compensatory equalization function. The top row uses 4 bits and replicates the results plotted in the fourth row of Figure \ref{fig:constant_slope_sims}. In subsequent rows, we increase the number of re-quantization bits. For each additional bit, we reduce $k$ (see Figure \ref{fig:requantization}) by 1, meaning we are adding least significant bits. For each trial $k=20-b_2$, where $b_2$ is the number of bits. We see only modest improvement in the spectral contamination as we increase the re-quantization bits from 4 to 8. The contamination exceeds 1 part in $10^5$ for all cases explored here.

If the number of bits could be increased sufficiently to accommodate the full dynamic range of the filter bank output, across all frequencies, we would no longer require equalization of the digital signal. For the OVRO-LWA, which has significant spectrum variation across its wide frequency range, this would require a large increase in the data rate from the re-quantizer to the correlator.

\subsection{Increase the Gain Prior to Equalization} \label{sec:increase_gain}

\begin{figure}
    \centering
    \includegraphics[width=\linewidth]{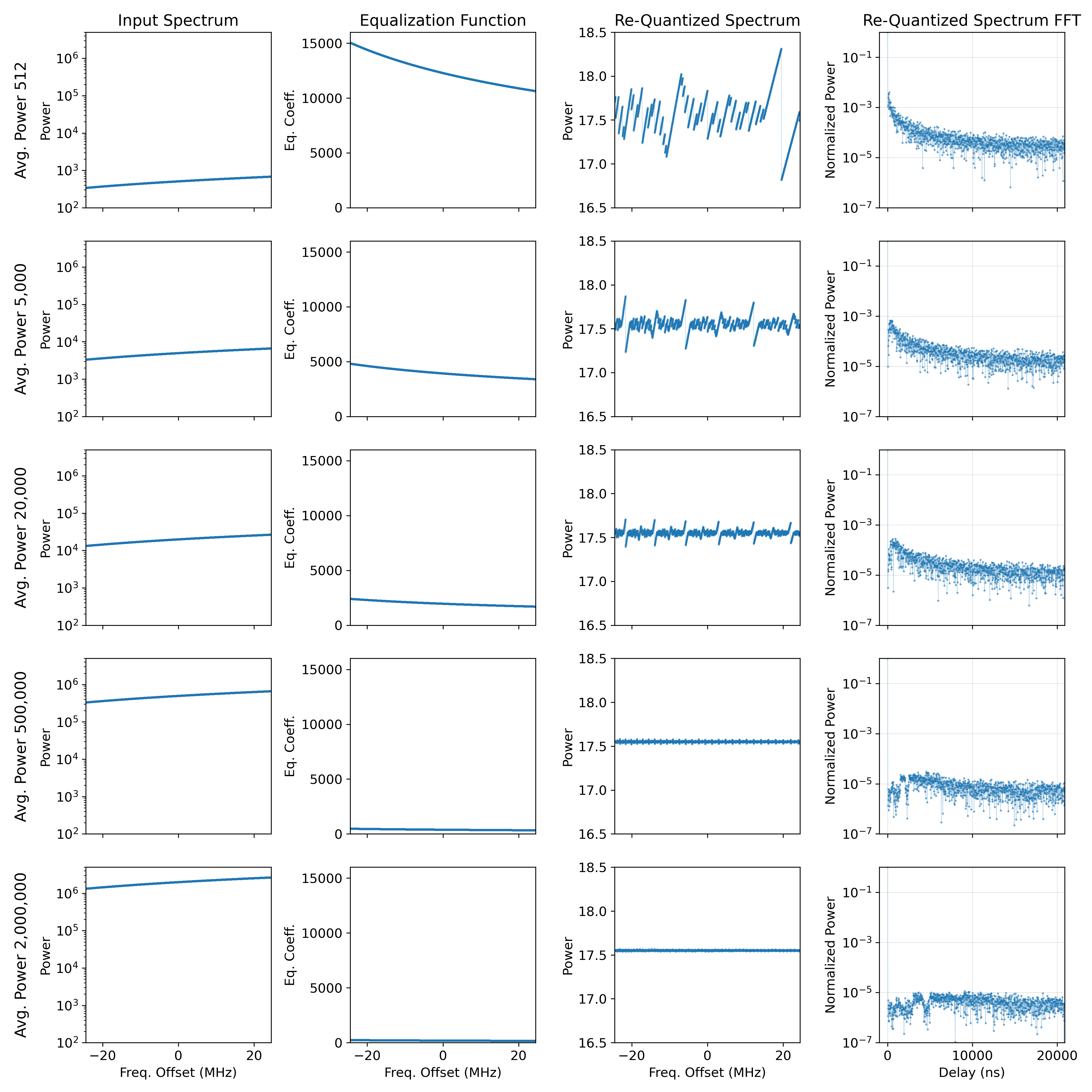}
    \caption{Simulated signals with varying average signal power. Each signal varies linearly across frequency and has a fractional variation of 100\% across 48.97 MHz. The top row has an average signal power of 512 and is equivalent to the fourth row in Figure \ref{fig:constant_slope_sims}. Subsequent rows increase the average signal standard deviation to 5,000, 20,000, 500,000, and 2,000,000. Increasing the signal power, and reducing the equalization function commensurately, reduces the systematic effect.}
\label{fig:power_var_sims}
\end{figure}

Increasing the pre-equalization gain makes the signal more finely quantized and causes more unique numbers to contribute to each re-quantized value (see the histogram in Figure \ref{fig:requant_hist}). This reduces the relative contribution of values on the bin boundaries and has a strong impact on mitigating spectral distortion. By placing more gain ahead of the equalizer ($g_a\,g_d\,g_f$ in Figure \ref{fig:flowchart}) and less after that point ($g_e\,g_q$), the size of the discontinuities can in principle be made as small as desired. Figure \ref{fig:power_var_sims} plots signals with varying power, using mean variances of 2,500, 10,000, 250,000, and 1,000,000. The increasing signal power drastically reduces spectral contamination. 

In practice, however, this is subject to constraints, as discussed in \S\ref{sec:overview}. If the analog gain $g_a$ is too large, the digitizer can saturate frequently. After digitization, if the gain is too large at any stage, the resulting numbers can overflow their finite-bit-width representations. Even within these constraints, gain adjustments in the OVRO-LWA signal path could increase the pre-equalization signal amplitude. This would reduce systematic spectral distortion, though not to the level required for 21 cm cosmology. It would also lead to higher fractional data loss, as an increased occurrence of overflow events would require flagging and data removal.

The simulations presented in Figure \ref{fig:power_var_sims} demonstrate that, even in the case that the simulated frequency discontinuities are very small compared to the mean signal power, there can be non-negligible spectral contamination apparent in the Fourier transformed signal, with the peak-normalized signal approaching or exceeding $10^{-5}$ for some delays. This shows that, for particular tunings of gains along the signal path, spectral effects from equalization may not be immediately apparent in the data. The semi-analytic simulations presented here represent the limit of infinite accumulation time and are noise-free. For the OVRO-LWA, these systematic effects are easily identifiable in the spectra measured from just seconds of data (see Figure \ref{fig:autocorr_spec}), but this may not be the case for all instruments. The effect could be swamped by thermal noise for short time integrations but nonetheless appreciable for the long integrations required by 21 cm analyses.

\subsection{Model and Correct the Effect}

The simulation presented in \S\ref{sec:simulation} is exact (assuming the underlying assumptions are correct) and deterministic. This means that, in principle, data could be corrected after the fact to undo the effect of equalization and re-quantization.


Further work is needed to determine with what precision the systematic effects explored in this paper can be retroactively corrected. The simulation approach presented in \S\ref{sec:simulation} represents only the signal processing stages after the filter bank, and assumes that the filter bank output can be represented by a quantized Gaussian random signal. It is not clear to what extent this assumption is valid. A more accurate simulation would include the full signal path upstream from the filter bank and capture quantization effects from the signal's initial digitization.

\subsection{Dithering}

Dithering is the process of intentionally adding noise in order to reduce systematic errors. It is used in many digital applications (see, for example, \citealt{schuchman1964, zames1976,delotto1986,wagdy1989,wagdyEffectVariousDither1989,lipshitz1992,wagdy1994,wannamaker2000}; and \citealt{gustafsson2013} for general treatments). \citealt{Offringa2016} presents an application of dithering to radio astronomy, specifically for error suppression in interferometric data compression. We show in simulation that dithering can effectively mitigate systematic spectral effects from equalization and re-quantization.

In this simulation, we add a small amount of random noise to the signal after equalization but before re-quantization. This causes equalized samples that would otherwise fall near the boundary of a re-quantization bin to have some probability of falling into the adjacent bin. As a result, the re-quantized signal no longer experiences sharp discontinuities as variations in equalization coefficient shift values across bin edges.

We generate dithering noise as samples from a Gaussian distribution that are independent for each time step. We choose a standard deviation equal to 10\% of the least-significant bit (LSB) of the re-quantized value (bit $k$ in Figure \ref{fig:requantization}).

\begin{figure}
    \centering
    \includegraphics[width=\linewidth]{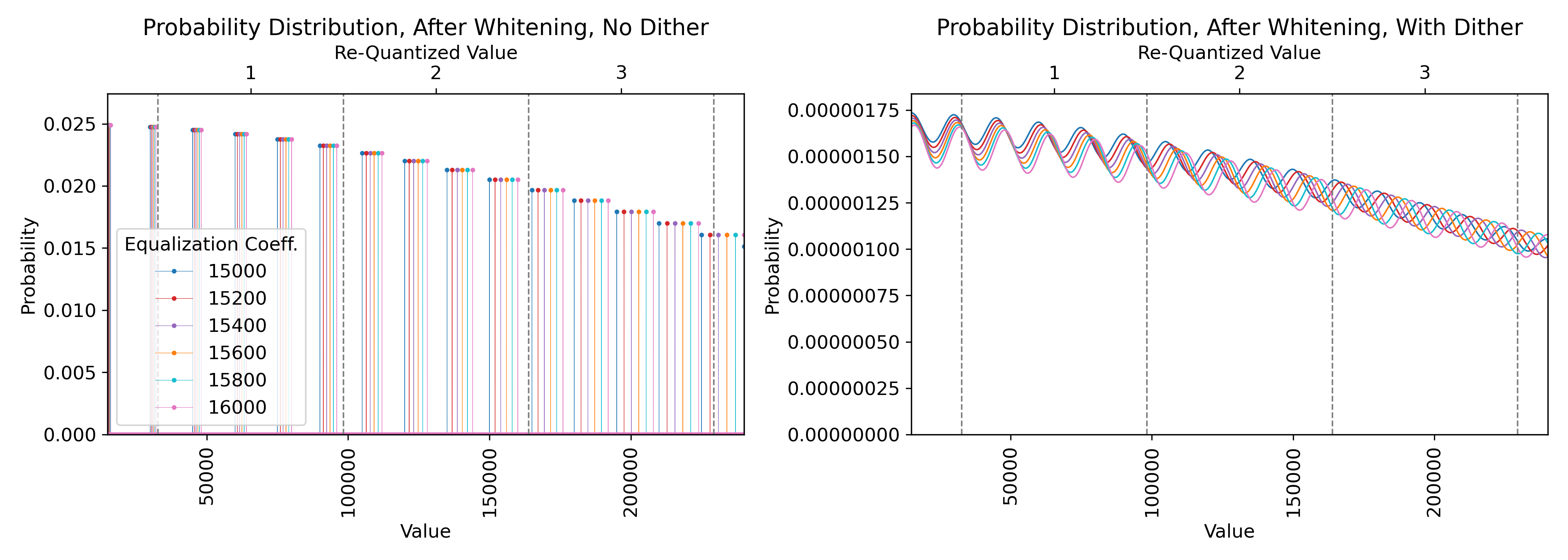}
    \caption{Simulated probability distribution of the equalized values before re-quantization, with and without dithering. The left panel does not include dithering and replicates the right panel from Figure \ref{fig:pdf_whitening}. It has a comb-like structure, as only multiples of the equalization coefficient have non-zero probability. The right panel represents the probability distribution with injected Gaussian noise (dither), using a dither standard deviation equivalent to 10\% of the re-quantization resolution. This corresponds to convolving the signal's probability distribution with that of the dithering noise. The result is that the sharp peaks in the left panel are smoothed. The vertical dashed lines denote boundaries of the re-quantization bins.}
\label{fig:pdf_dithering}
\end{figure}

\begin{figure}
    \centering
    \includegraphics[width=\linewidth]{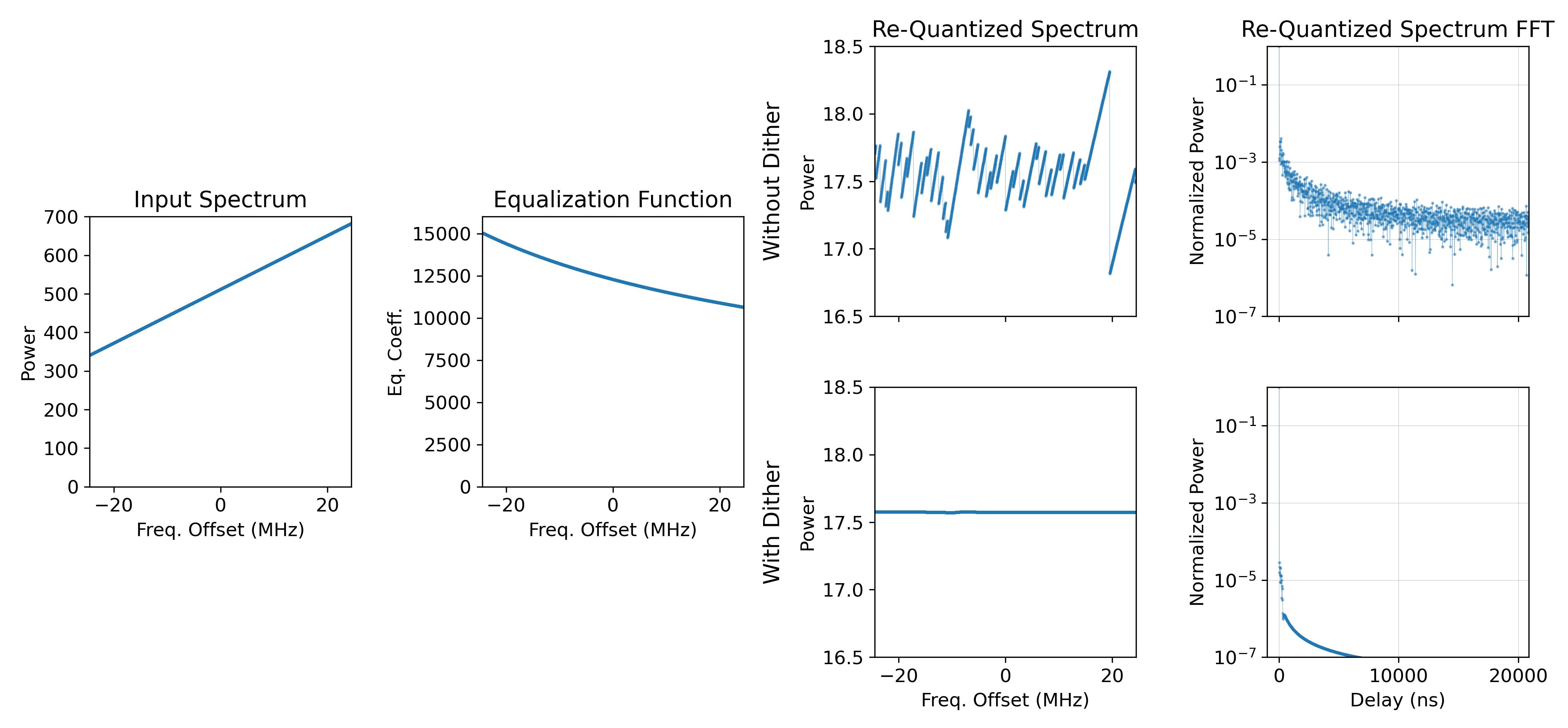}
    \caption{Simulated signals with and without dithering. Once again, the input signal and equalization function are equivalent to the fourth row in Figure \ref{fig:constant_slope_sims}. The top row corresponds to typical re-quantization; the bottom row introduces random Gaussian noise (dither) before re-quantization. For this simulation, we use a standard deviation equal to 10\% of the re-quantization resolution for the dither, although this could be adjusted. Dithering is extremely effective in reducing spectral contamination.}
\label{fig:dithering_sim}
\end{figure}

In our semi-analytic simulation, we compute the equalized signal's probability distribution with dithering by convolving its probability distribution without dithering (Figure \ref{fig:pdf_whitening}) with the probability distribution of the dithering noise. The result is plotted in Figure \ref{fig:pdf_dithering}. Propagating this probability density through the re-quantization step gives the expected 4-bit signal, plotted in Figure \ref{fig:dithering_sim}.

Dithering increases the additive noise, but so does re-quantization by adding to the quantization noise. If the LSB has value 1, then quantization noise has a variance of 1/12 \citep{widrow2010}. For dithering with a standard deviation 0.1 after re-quantization, or variance of 0.01, the dithering noise is about 1/8 of the quantization noise.

In a practical implementation, dithering would use a pseudo-random numbers rather than truly random numbers. These numbers should be independent for each time step. They need not be independent for each frequency channel of the same signal, but they must be very nearly independent among different signals to avoid bias when signals are cross-correlated.


\section{Conclusion}
\label{sec:conclusion}

We have shown that digital equalization (``whitening'') and re-quantization of radio astronomical signals can cause systematic spectral distortion. This appears as discontinuous frequency structure and occurs even when the original signal and the equalization function are both spectrally smooth. While this may not be important many science applications, especially when mitigated in calibration, it is highly relevant for applications with extreme requirements for spectral fidelity, such as 21 cm cosmology.

There are multiple effective mitigation strategies for this distortion. It can be fully avoided if digital equalization can be made unnecessary by retaining sufficient sample bits in every frequency channel to represent the full dynamic range of the signal. In some cases, analog whitening filters may reduce the dynamic range enough to make this practical. Piecewise-constant equalization functions eliminate the effect across corresponding segments of the total bandwidth. We show in simulation that increasing the signal power prior to equalization and dithering are extremely effective for reducing the distortion; modest increases in the number of re-quantization bits are minimally effective.

The magnitude of the distortion is strongly dependent on the details of the instrument's signal processing---the distribution of gains along the signal path, the number of bits used at each stage, and the pre-equalization spectrum. For the OVRO-LWA, low pre-equalization gain and large spectral dynamic range produce extremely strong systematic spectral effects that are readily identifiable in the spectra of individual signals. For other instruments with a similar processing architecture, the effect may be more subtle if the pre-equalization signal is more finely quantized or the spectrum is closer to flat. In some cases, it may be indistinguishable from random noise for short time integrations but still occur at a level that contaminates 21 cm cosmology measurements. 

This effect should be fully understood for any instrument that uses digital equalization and re-quantization and aims to perform measurements with good spectral fidelity. We show that it can be well quantified through simulations such as those used here. Mitigation strategies should be implemented if the systematic exceeds allowable tolerances (relative spectral distortion of 1 part in $10^5$ for 21 cm cosmology applications).

\ack{We thank Jack Hickish and Real-Time Radio Systems Ltd for helpful discussions and technical support.}

\funding{R.B.\ is supported by the National Science Foundation Award No.\ 2303952.}


\data{The data that support the findings of this study are available upon reasonable request from the authors.}


\bibliography{sample701}{}
\bibliographystyle{aasjournalv7}

\end{document}